\documentclass[%
 reprint,
 amsmath,amssymb,
 aps,
]{revtex4-1}

\usepackage{graphicx}
\usepackage{subcaption}
\usepackage{dcolumn}
\usepackage{bm}
\usepackage[justification=justified,singlelinecheck=false]{caption}

\begin{document}
\captionsetup{justification=justified,singlelinecheck=false}
\preprint{APS/123-QED}

\title{Opportunities With Decay-At-Rest Neutrinos From Decay-In-Flight Neutrino Beams}

\author{Christopher Grant}
 \email{cpgrant@bu.edu}
 \affiliation{Department of Physics, Boston University, Boston, MA, 02215, USA}

\author{Bryce Littlejohn}%
 \email{blittlej@iit.edu}
\affiliation{Physics Department, Illinois Institute of Technology, Chicago, IL 60616, USA}%

\date{\today}

\begin{abstract}
Neutrino beam facilities, like spallation neutron facilities, produce copious quantities of neutrinos from the decay at rest of mesons and muons.  
The viability of decay-in-flight neutrino beams as sites for decay-at-rest neutrino studies has been investigated by calculating expected low-energy neutrino fluxes from the existing Fermilab NuMI beam facility.  
Decay-at-rest neutrino production in NuMI is found to be roughly equivalent per megawatt to that of spallation facilities, and is concentrated in the facility's target hall and beam stop regions.
  Interaction rates in 5 and 60~ton liquid argon detectors at a variety of existing and hypothetical locations along the beamline are found to be comparable to the largest existing decay-at-rest datasets for some channels.  
The physics implications and experimental challenges of such a measurement are discussed, along with prospects for measurements at targeted facilities along a future Fermilab long-baseline neutrino beam.


\end{abstract}

\maketitle

\section{Introduction}

Neutrino beams have played an integral role in the understanding of neutrino oscillations and interactions in matter.  
Beam neutrinos are produced in the decay of $K$ and $\pi$ meson secondaries from collisions of an accelerated proton beam with a target.  Using focusing magnetic horns, mesons can be directed to and decayed in a drift region, producing a tuneable beam of high-energy neutrinos.  
Recently, the  J-PARC Neutrino Experimental Facility~\cite{Abe:2012av}, the CNGS beam at CERN~\cite{Giacomelli:2007df}, and the Booster~\cite{AguilarArevalo:2008yp} and Main Injector~\cite{arXiv.1507.06690} beams at Fermilab have served as primary neutrino physics facilities producing decay-in-flight (DIF) neutrinos and antineutrinos in the $\sim$0.5-50~GeV range.  
Detectors situated in the paths of these beams 
have produced an impressive array of physics including measurement of neutrino oscillations~\cite{Abe:2015awa, Adamson:2014vgd, Agafonova:2015jxn}, searches for sterile neutrinos~\cite{AguilarArevalo:2009xn}, and new insights into nuclear/nucleon structure~\cite{Garvey:2014exa}.

While producing a high-energy neutrino beam, these facilities also produce a far-more-copious flux of lower-energy neutrinos from the decay at rest (DAR) of pions, kaons, and muons.  
A large majority of charged mesons produced by proton beams are not focused by the horns and instead DAR in or near the target.  
Of the horn-focused component, many mesons and product muons reach the end of the beam drift region without decaying, instead terminating via DAR in large beam stops.  
The resultant fluxes of DAR $\nu_e$, $\nu_{\mu}$, and $\overline{\nu}_{\mu}$ are similar in character to those produced at nuclear physics facilities primarily dedicated to the production of spallation neutrons~\cite{Louis:1997bs}.  
These neutrinos are useful for a variety of purposes.  
The low energies and well-understood neutrino flavor composition and spectra of kaon, pion, and muon DAR allow for sterile oscillation searches in a variety of channels at short baselines~\cite{Aguilar:2001ty, PhysRevD.85.013017, MLF:2013, OscSNS:2013, Axani:2015dha}.  
Pion and muon DAR neutrinos in the 10-50~MeV are also ideal for detection via neutrino-nucleus coherent scattering~\cite{COHERENT:2013}.  
Finally, observation of $\mu^{+}$ DAR $\nu_e$ in a liquid argon TPC (LArTPC) would elucidate the range of complex final states produced by supernova and solar neutrinos in future large underground LArTPCs, such as DUNE~\cite{Adams:2013qkq}.

Due to their high power and protons-on-target (POT), existing spallation neutron facilities have been the primary focus of completed or proposed DAR experiments.
Less attention has been given in the literature to fluxes and facilities available for decay-at-rest experiments at decay-in-flight neutrino beam facilities.  
Existing studies have focused on the Fermilab-based NuMI and BNB beams: some have indicated promising physics capabilities with respect to coherent neutrino scattering~\cite{PhysRevD.89.072004}, while others have highlighted the advantages of spallation neutron source-based experiments relative to  Fermilab-based opportunities~\cite{PhysRevD.89.073007}.  

In this paper, we seek to clarify the potential of current and future DAR opportunities at Fermilab by more broadly exploring the low-energy neutrino flux provided by NuMI neutrino beamline.  
We present simulated fluxes of pion, muon and kaon DAR neutrinos separately for the NuMI target and beam dump and give expected fluxes and interaction rates in existing Fermilab LArTPCs and in hypothetical LArTPC deployments in or near existing Fermilab facilities.  
We then discuss the physics potential of these scenarios, some of which may be achievable with existing detectors and modest extensions of existing Fermilab infrastructure.  
Similar opportunities with future Fermilab neutrino beams are also discussed.  

\section{Method}

The calculation of neutrino production by the NuMI beam was performed using the NUMIX collaboration's G4NuMI software, a pure \textsc{Geant4}-based code~\cite{Agostinelli2003250,1610988} that propagates particles through a realistic geometry of the NuMI beam facility.  The geometry includes the target hall containing a graphite target followed by two magnetic focusing horns, all of which is encased in layers of steel and concrete shielding.  A 675~m long, 2~m diameter steel decay pipe begins at the end of the target hall and is filled with helium gas.  A beam dump, called the NuMI hadron absorber, is located at the end of the decay pipe and is made up of a 1.3 m wide $\times$ 1.3 m high $\times$ 4.75 m long core containing aluminum blocks and ten steel blocks, surrounded by additional steel and concrete.  
The entire beam line is centered along one axis in a cylindrical world volume of rock 1200 m long and 25 m in diameter extending roughly 400 m past the absorber end.  
The origin in Cartesian coordinates, $(x,y,z)$=$(0,0,0)$~m, is defined to be the center of the upstream face of the first horn.  The beam of 120 GeV protons is generated in the $+$z-direction at $(0,0,-4)$~m.  Further description of the NuMI facility can be found in Ref.~\cite{arXiv.1507.06690}.


The main source of DIF neutrino production comes from $K^{+} \rightarrow \nu_{\mu}\mu^{+}$ and $\pi^{+} \rightarrow \nu_{\mu}\mu^{+}$, with decay products boosted in the +$z$ direction.  
%
%
For DAR, the primary neutrino production channels are $K^{+}\rightarrow\nu_{\mu}\mu^{+}$, $\pi^{+}\rightarrow\nu_{\mu}\mu^{+}$, and $\mu^{+}\rightarrow\overline{\nu}_{\mu}\nu_{e}e^{+}$. Other contributions also come from $\mu^{-}p \rightarrow n\nu_{\mu}$ (nuclear capture) and $\mu^{-} \rightarrow \nu_{\mu}\overline{\nu}_{e}e^{-}$ (decay in orbit around the atomic nucleus).  Fluxes incident on 1 m$^{2}$ areas are calculated at pre-defined locations in $(x,y,z)$ coordinates.  To good approximation, all DAR neutrinos are emitted isotropically.  

The existing G4NuMI beam simulation can be tuned to stop tracking particles that fall below a user-defined kinetic energy.  
This adjustment is useful for flux calculations of DIF neutrinos above several hundred MeV, decreasing computation time dramatically.  
However, tracking cuts on the kinetic energy did not allow production of DAR neutrinos and were not reliable below $\sim$250 MeV.  
In this work the cut was removed and daughter particles were tracked down to zero kinetic energy.  
Flux calculations of DIF neutrinos at MINOS were reproduced to confirm that no significant alteration to fluxes above 250~MeV occurred as a result.

\section{Results} 

We first present absolute rates of DAR neutrino production along the NuMI DIF beamline.  
The rates of the most significant contributions are found to be 0.5 $K^{+}$, 11.0 $\pi^{+}$, 27.3 $\mu^{+}$, and 1.9 $\mu^{-}$ DAR $\nu$/proton.
These totals are a factor of $\sim$200-400 higher per proton than lower-energy spallation $\nu$ sources~\cite{OscSNS:2013,Axani:2015dha}, compensating for the lower overall protons on target produced in DIF facilities.   
When production is compared per MW beam power we find DAR production at DIF facilities to be roughly equal to spallation facilities.  
We also note that this same general equivalence per MW also applies between the 700~kW NuMI and 32~kW Booster beamlines, indicating roughly an order of magnitude lower DAR neutrino production at the Booster.

\begin{figure}[t]
\centering
\begin{subfigure}{0.8\columnwidth}
\centering
   \includegraphics[trim={2mm 0mm 2mm 8mm},clip,width=1.0\textwidth]{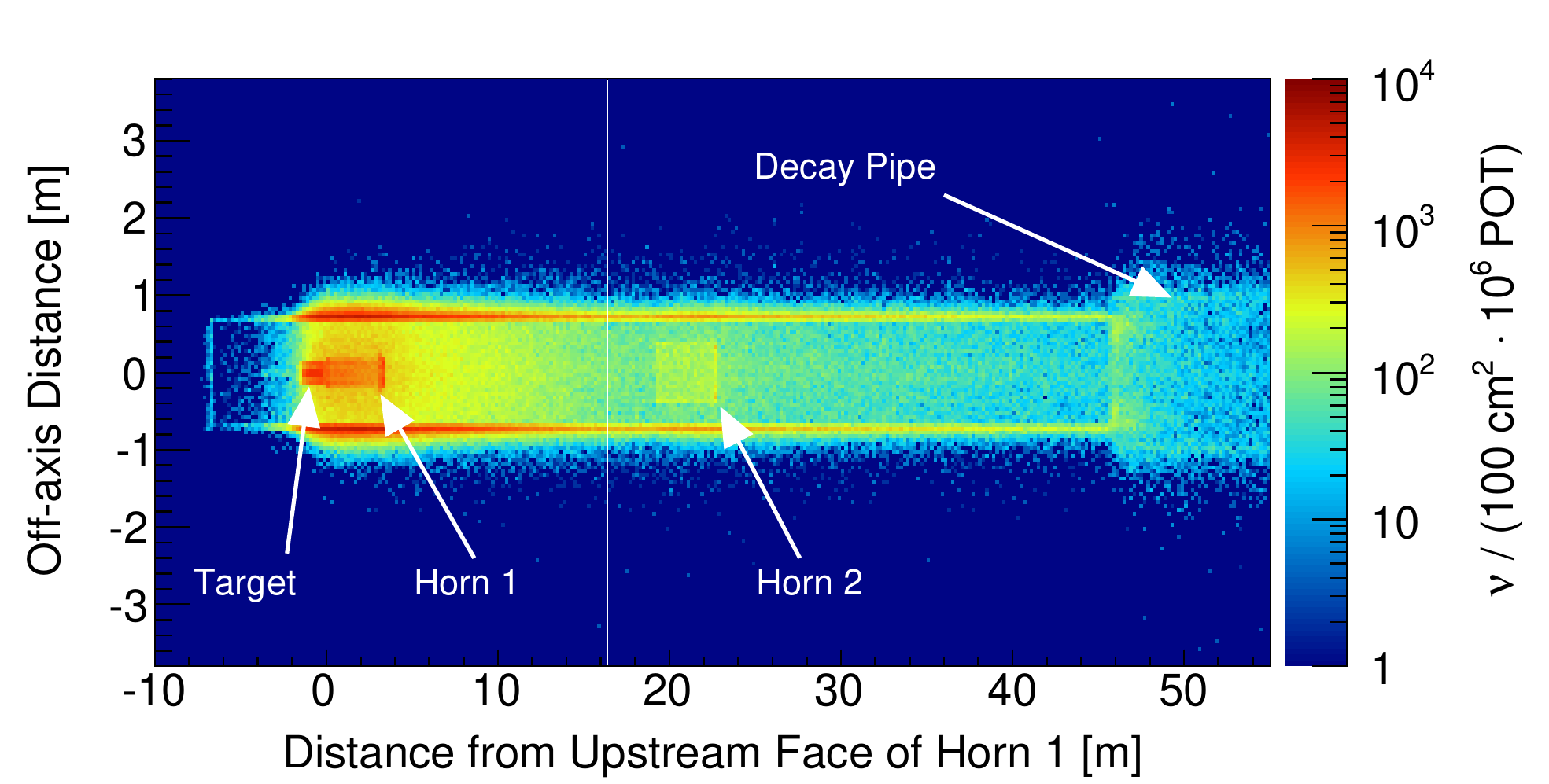}
\end{subfigure}
\begin{subfigure}{0.8\columnwidth}
\centering
  \includegraphics[trim={2mm 0mm 2mm 6mm},clip,width=1.0\textwidth]{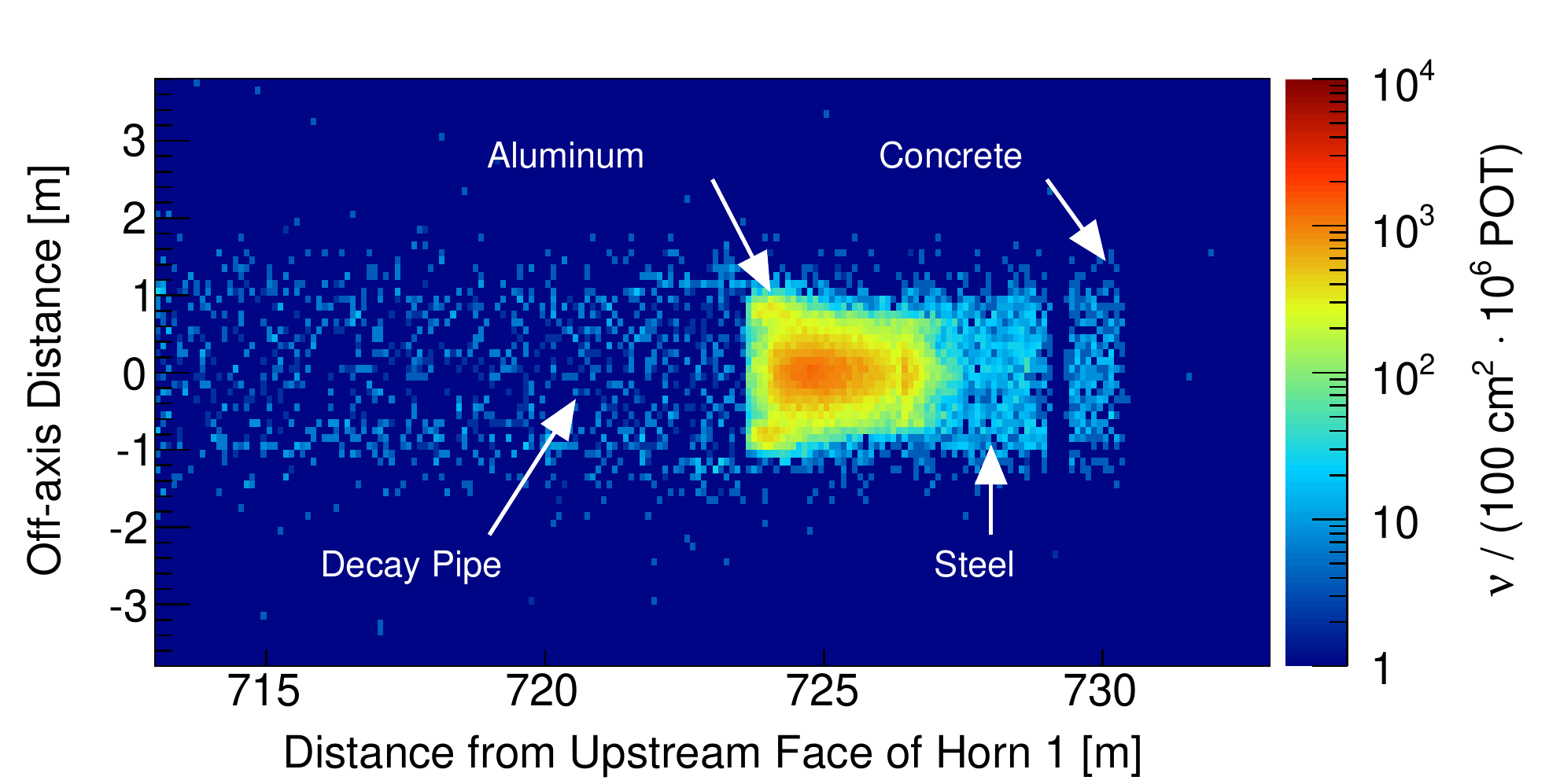}
\end{subfigure}
\vspace{-1mm}
\caption[]{\raggedright Neutrino origin distributions inside the NuMI target hall (top) and hadron absorber (bottom), as viewed from above.}
\label{fig:nu_production}
\vspace{-4mm}
\end{figure}

The flux from DAR neutrinos is highest in two locations:  the target hall and the hadron absorber.  
This can be seen in Fig.~\ref{fig:nu_production} where the distribution of stopped neutrino parents is shown in the 2D plane, viewed from above. 
Neutrino production is highest in the target hall:  
80\% of $\nu$ are produced within a 2~m x 2~m x 60~m rectangular box surrounding these target and horn areas.   
Of this majority, only 4\% and 6\% are produced directly within the target and horns, respectively, highlighting the important role of surrounding shielding material in low-energy neutrino production.  
On the other end of the decay pipe, a smaller but substantial 12\% of DAR flux is produced within a 3~m x 3~m x 4~m rectangular box encompassing the hadron absorber core.


%

\begin{figure}[b]
	\begin{center}
		\includegraphics[trim={0 0 19mm 12mm},clip,width=0.9\columnwidth]{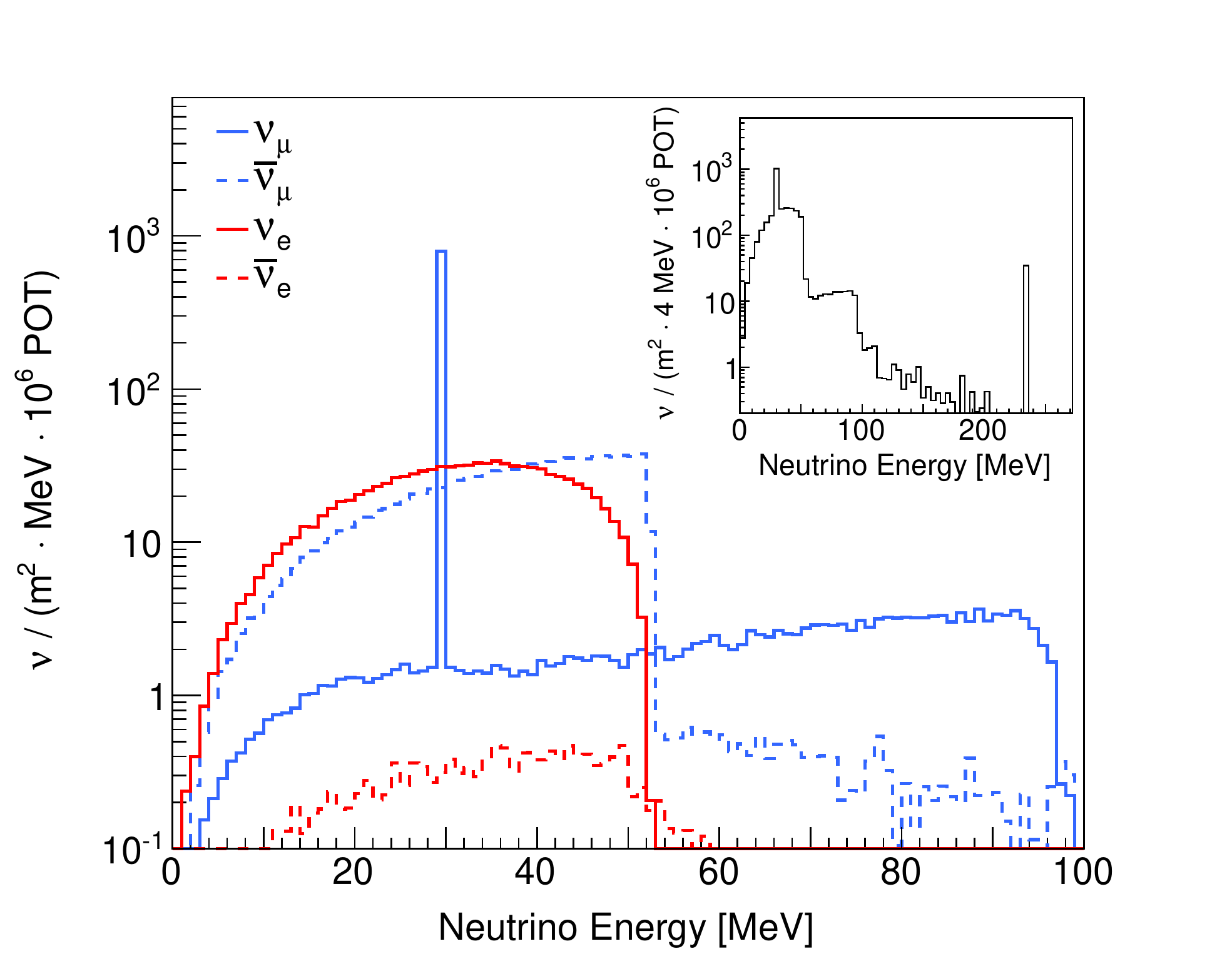}
		\vspace{-0.6cm}
	\end{center}
	\caption[]{\raggedright Neutrino flux at 25 m distance from the NuMI target with an inset showing the total flux for all flavors.}
	\label{fig:nu_flux_target}
\end{figure}

The neutrino flux below 100 MeV is shown in Fig.~\ref{fig:nu_flux_target} at an example distance of 25 m off-axis directly to the side of the target.  
The flux below 53 MeV is almost entirely from $\pi^{+}$ and $\mu^{+}$ DAR with minor (4\%) DIF contamination.  The two-body $\pi^{+}$ decay produces a mono-energetic 29.8 MeV $\nu_{\mu}$, and the three-body $\mu^{+}$ decay produces a continuous spectrum of $\nu_{e}$ and $\overline{\nu}_{\mu}$ up to 53 MeV. The shoulder extending up to 100 MeV is a byproduct of $\mu^{-}$ capture on the nuclei of material in the target hall.  The inset of Fig.~\ref{fig:nu_flux_target} shows the total flux from all flavors out to 250 MeV to make visible the flux of mono-energetic 235.5 MeV $\nu_{\mu}$ from the two-body $K^{+}$ DAR.  DIF contamination to $K^{+}$ DAR, 1\% between 230 and 240 MeV, is also low.  Spectral features from the absorber are similar, with 2\% DIF contamination below 53~MeV and a higher 13\% for $K^{+}$ DAR between 230 and 240~MeV.  

\begin{table*}[t!]
\caption{\raggedright Flux values per $6 \times 10^{20}$ POT at four locations near the NuMI beam line.  The $\mu^{+}$ and $\mu^{-}$ DAR flux values are integrated over an energy range of 0 to 53 MeV. }
\vspace{-3mm}
\centering
\setlength\extrarowheight{1mm}
\begin{tabular}{l@{\hskip 1.0cm} c@{\hskip 0.7cm} c@{\hskip 0.3cm} c@{\hskip 0.7cm} c@{\hskip 0.3cm} c@{\hskip 0.7cm} c}
\hline
\hline
Location\rule{0pt}{1.2\normalbaselineskip} & $\pi^{+}$ DAR & \multicolumn{2}{l}{\hspace{0.9cm}$\mu^{+}$ DAR} & \multicolumn{2}{l}{\hspace{0.9cm}$\mu^{-}$ DAR} & $K^{+}$ DAR\\
$(x,y,z)$ [m] & $\Phi_{\nu_{\mu}}$ [m$^{-2}$] & $\Phi_{\nu_{e}}$ [m$^{-2}$] & $\Phi_{\overline{\nu}_{\mu}}$ [m$^{-2}$] & $\Phi_{\nu_{\mu}}$ [m$^{-2}$] & $\Phi_{\overline{\nu}_{e}}$ [m$^{-2}$] & $\Phi_{\nu_{\mu}}$ [m$^{-2}$] \vspace{1mm}\\
\hline
\rule{0pt}{1.2\normalbaselineskip}$(15,20,0) \rightarrow $ 25 m from target      & 4.8$\times 10^{17}$     & 5.8$\times 10^{17}$ & 5.8$\times 10^{17}$ & 1.3$\times 10^{16}$ & 8.0$\times 10^{15}$ & 2.1$\times 10^{16}$ \\

$(0,40,0) \rightarrow $ 40 m from target      & 2.0$\times 10^{17}$     & 2.5$\times 10^{17}$ & 2.5$\times 10^{17}$ & 5.8$\times 10^{15}$ & 3.5$\times 10^{15}$ & 9.4$\times 10^{15}$ \\

$(15,-1,753) \rightarrow $ 32 m from absorber    & 4.6$\times 10^{16}$     & 5.8$\times 10^{16}$ & 5.8$\times 10^{16}$ & 4.6$\times 10^{14}$ & 3.9$\times 10^{14}$ & 1.8$\times 10^{15}$\\

$(53,76,679) \rightarrow$ $\mu$BooNE \vspace{1mm}       & 6.5$\times 10^{15}$  & 7.8$\times 10^{15}$ & 7.8$\times 10^{15}$ & 1.5$\times 10^{14}$ & 1.2$\times 10^{14}$ & 2.6$\times 10^{14}$ \\
\hline
\hline
\end{tabular}
\label{tab:flux_position}
\end{table*}

The DAR neutrino flux produced by the $6 \times 10^{20}$ POT from one year of upgraded 700~kW NuMI beamline operation was calculated for four $(x,y,z)$ positions off-axis of the beam line, and is reported in Table~\ref{tab:flux_position}.  
Positions $(15, 20, 0)$ and $(0,40,0)$ are located near the target: the former location is underground, adjacent to the existing NuMI target hall shaft, while the latter is directly above the the target hall in un-improved space at ground level.  
Position $(15,-1,753)$ is 15~m off-axis and 28~m downstream from the hadron absorber in a large, serviced underground off-axis tunnel connecting the NuMI decay pipe and near detector hall.  
Finally, $(53,76,679)$ is the location of the MicroBooNE experiment.  

For further illustration, charged and neutral current neutrino interaction rates in liquid argon detector target material were then calculated at the examined locations.  
Detector target sizes of 5~t were chosen for the first three locations to mirror the size of small LArTPCs currently under construction~\cite{arXiv.1309.1740}, while a 60~t fiducial volume was placed at the MicroBooNE detector location~\cite{Antonello:2015lea}.  
For energies less than 53 MeV, theoretical calculations of the relevant cross-sections were utilized~\cite{1742-6596-308-1-012012}, while interaction rates from the 235.5 MeV $\nu_{\mu}$ flux was calculated using the charged-current quasi-elastic cross-section from Ref.~\cite{PhysRevC.86.015505}.  
Resultant interaction rates are given in Table~\ref{tab:rate_position}.  


\begin{table*}[t!]
\caption{\raggedright Interactions per $6 \times 10^{20}$ POT at four locations near the NuMI beam line.  All rates are integrated over an energy range of 0 to 53 MeV with the exception of the $K^{+}$ DAR 235.5 MeV $\nu_{\mu}$ interaction rate. }
\vspace{-3mm}
\centering
\setlength\extrarowheight{1mm}
\begin{tabular}{l@{\hskip 0.7cm} c@{\hskip 0.7cm} c@{\hskip 0.5cm} c@{\hskip 0.5cm} c@{\hskip 0.5cm} c@{\hskip 0.6cm} c}
\hline
\hline
Location\rule{0pt}{1.2\normalbaselineskip} & & & & & & \\
$(x,y,z)$ [m] & LAr Mass [t] & CC $\nu_{e}$ & CC $\overline{\nu}_{e}$ & NC $\nu_{\mu}$ & NC $\overline{\nu}_{\mu}$ & CC $\nu_{\mu}$ (235.5 MeV) \vspace{1mm}\\
\hline
\rule{0pt}{1.2\normalbaselineskip}$(15,20,0) \rightarrow $ 25 m from target & 5.0 & 1.1$\times 10^{3}$ & 1.4 & 1.2$\times 10^{2}$ & 4.1$\times 10^{2}$ & 4.5$\times 10^{3}$ \\

$(0,40,0) \rightarrow $ 40 m from target & 5.0 & 4.9$\times 10^{2}$ & 0.6 & 5.0$\times 10^{1}$ & 1.8$\times 10^{2}$ & 2.0$\times 10^{3}$ \\

$(15,-1,753) \rightarrow $ 32 m from absorber & 5.0 & 1.1$\times 10^{2}$ & 0.1 & 1.0$\times 10^{1}$ & 4.0$\times 10^{2}$ & 4.0$\times 10^{2}$ \\

$(53,76,679) \rightarrow$ $\mu$BooNE \vspace{1mm} & 60.4 & 1.8$\times 10^{2}$  & 0.3 & 1.9$\times 10^{1}$ & 6.5$\times 10^{1}$ & 6.9$\times 10^{2}$\\
\hline
\hline
\end{tabular}
\label{tab:rate_position}
\end{table*}

Statistics for the various neutrino types are sizeable in spite of the small detector sizes utilized.  
For comparison, a $\nu_e$ sample from two years of data-taking at either examined target location could eclipse the largest existing datasets of this type from KARMEN and LSND~\cite{KARMEN_Thesis, Auerbach:2001hz}, depending on the achieved detection efficiency of the utilized detector.   
Similar statistics could be achieved near the NuMI absorber with a slightly larger LArTPC over a multi-year run period.  
We emphasize that relatively modest improvements would be required to host a 5~t LArTPC at some of the examined locations.  
We also note again that LArTPCs of similar magnitude are presently under construction or design, with many slated for installation within the next five years at Fermilab~\cite{arXiv.1309.1740, Antonello:2015lea}.  While significant experimental challenges will be present for these detector-location combinations, the impressive DAR $\nu$ event rates from DIF sources clearly demand further attention.  

\subsection{Physics Opportunities}\label{physopp} 
A number of physics opportunities of significant current interest can be pursued utilizing the DAR $\nu$ produced by DIF facilities.  We make no attempt to quantify experimental sensitivities to the physics discussed below, however, a significant DAR $\nu$ flux warrants more extensive investigation into any one of these possibilities.

The similarity in energy between $\mu^{+}$ DAR $\nu_e$ and supernova-produced $\nu_e$ presents a clear opportunity for understanding supernova neutrino detection in the DUNE LArTPC~\cite{Adams:2013qkq}.  
Unlike the inverse beta decay cross-section of $\overline{\nu}_{e}$ on hydrogen, which is known to better than 0.2$\%$ and has a single final state, the charged-current (CC) reaction $^{40}$Ar$(\nu_{e},e^{-})^{40}$K$^{*}$ has many inherent complexities:
\begin{enumerate}
\item The ground state ($0^{+}$) to ground state ($4^{-}$) transition is third-forbidden, while thresholds and transition strengths for the $\sim$20 allowed and forbidden excited states are poorly constrained~\cite{PhysRevC.58.3677,PhysRevC.80.055501}.
\item De-excitation gammas from over half of the known excited states have not been measured.
\item A nucleon evaporation threshold of $\sim$12~MeV indicates production of free final-state nucleons, which will have a major yet un-characterized impact on reconstruction of $\nu$ energies.
\end{enumerate}
\noindent Predictions of the total CC $\nu_{e}$ cross-section obtained from measurements of $^{40}$Ti beta decay, $^{40}$Ar$(p,n)^{40}$K$^{*}$ reactions, and theoretical calculations from QRPA models differ by more than 30\% at an incident neutrino energy of 12 MeV. Similar uncertainties will also affect interpretation of the neutral current (NC) reaction $^{40}$Ar$({\nu},\nu')^{40}$Ar$^{*}$~\cite{PhysRevC.73.054306, arXiv.1503.08095}.  
The NuMI-produced low-energy $\nu$ fluxes described here would allow direct study of neutrino-argon interaction final states in ton-scale LArTPCs.  
Such a measurement would facilitate realistic, data-driven design requirements for TPC, photon detection, and trigger systems to ensure high-efficiency supernova neutrino detection in large underground LArTPCs.

Both the target and absorber region represent neutrino sources that could be used for short-baseline oscillation studies.  
  The baseline and energy ranges of fluxes in the locations examined above would provide sensitivity in the eV-scale neutrino mass-splitting region suggested by existing global fits~\cite{Kopp:2013}.  
  Fermilab's DIF facilities outlined above above offer a few unique advantages for such a measurement.  
  The sizable interaction rates of both 235.5~MeV $K^{+}$-produced $\nu_{\mu}$ and $\mu^{+}$-produced $\nu_e$ could allow for production of simultaneous searches in multiple disappearance channels.  
 In addition, the examined locations at ground level and in the off-axis NuMI tunnel may naturally facilitate deployment at multiple baselines, allowing allowing for more precise oscillation searches via relative rate and spectrum comparisons.  
Beyond this, absolute 235.5 $K^{+}$ DAR $\nu_{\mu}$ detection rate measurements can also provide valuable constraints on cross-sections, as described in Ref.~\cite{PhysRevD.89.073007}, and on meson production yields in DIF facilities.

  The discussion of fluxes and physics opportunities has thus far been constrained to existing Fermilab beam facilities.  
  It should also be mentioned that the design of future DIF facilities at Fermilab, in particular the Long Baseline Neutrino Facility (LBNF), is presently underway.  
Current LBNF beamline designs call for more than a three-fold increase in power producing a similar increase in total DAR fluxes.  
Moreover, current timelines for this facility's construction are amenable to timely identification and development of a dedicated facility for DAR $\nu$ detection within proposed target or absorber facilities.  
Such a facility would provide enhanced ability to address the physics goals described above, enabling DUNE to properly interpret both supernova and long-baseline $\nu_e$ signatures~\cite{Gandhi:2015} and to constrain dominant background and systematics contributions.  
Similar flux studies utilizing LBNF beamline simulations are currently underway to identify possible locations and beamline configurations  producing optimized DAR fluxes.

\subsection{Experimental Backgrounds}\label{exbg}  
To access the physics opportunities described above, relevant backgrounds at these energies must be mitigated.

Despite inherent suppression from the duty factor of Fermilab's DIF beams, cosmic-induced backgrounds may pose a challenge in the detection of $<$300 MeV neutrinos in LArTPCs near the surface, as outlined in Ref.~\cite{Antonello:2015lea}.  While these backgrounds will complicate measurements at MicroBooNE and the on-surface near-target location considered here, the other locations examined are protected by modest overburdens.  LArTPCs at the near-target $(15,20,0)$ and near-absorber $(15,-1,753)$ positions would sit beneath 20 and 85~m of rock, respectively, completely removing the primary hadronic cosmic ray component.  
Reduced muon-related backgrounds at these depths have been extensively studied in recent $\theta_{13}$ experiments.  
Reported muon fluxes of 5.11 and~1.16~m$^{-2}$s$^{-1}$ with 45 and~98~meters rock overburden (120 and 250 meters-water-equivalent)~\cite{An20158,1.4928019} indicate that sub-dominant cosmic backgrounds may be possible with LArTPCs at the examined locations, especially with implementation of external muon tagging systems and  topological discrimination of electron/gamma signals from cosmic muons and their spallation products.
We also note that a DAR facility within a future LBNF target or absorber hall would likely benefit from an even larger degree of overburden.

\begin{figure}[t]
	\begin{center}
		\includegraphics[trim={7mm 0 12mm 5mm},clip,width=0.95\columnwidth]{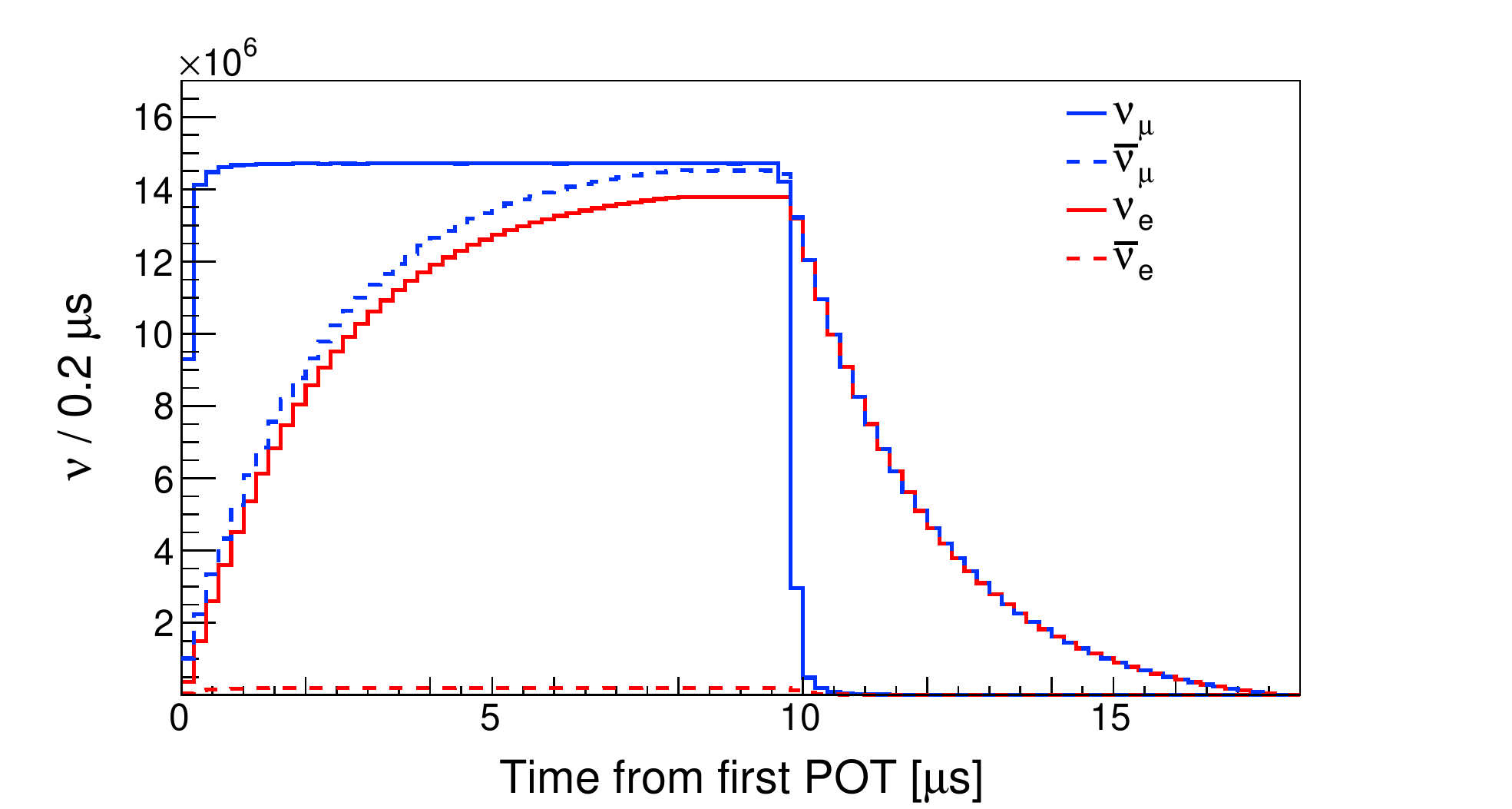}
		\vspace{-0.4cm}
	\end{center}
	\caption[]{\raggedright Time profile of neutrino production in the target hall during and after the 10 $\mu$s NuMI beam spill.}
	\label{fig:nu_time_target}
\end{figure}

For DAR pure-flavor oscillation or supernova studies, interactions of other neutrino flavors provide additional backgrounds to consider.  
These backgrounds may be addressed by utilizing the time-structure of neutrino interactions.  
The flavor composition near the target varies during a 10~$\mu$s NuMI spill due to differing meson and $\mu$ decay times, as shown in Fig.~\ref{fig:nu_time_target}.  
A similar structure is visible in the absorber, but is delayed by 2.3 $\mu$s due to the travel time down the decay pipe. 
A LArTPC utilizing sub-$\mu$s resolution from its light collection system can use this time structure to select specific neutrino flavors.  
As an example, for the $(15,20,0)$ location, by cutting all events in the first 10.4 $\mu$s of the beam spill, the total $\nu_{\mu}$ contribution to detected event rates below 53 MeV can be reduced by three orders of magnitude with less than one order of magnitude loss in $\nu_e$ signal statistics.  
It is likely that such analysis choices would only be statistically feasible at a NuMI-like DIF facility in a larger LArTPC, or in a future dedicated LBNF DAR facility.  
TPC topology and calorimetric cuts should naturally be studied further, as they are likely to provide more efficient separation of contributions from differing flavors and interaction channels.
It should be noted that, with respect to timing separation cuts, spallation facilities are likely to be more efficient due to their shorter beam spills.  Additionally, the delayed-coincidence tagging of $\overline{\nu}_{e}$ events in hydrocarbon-based scintillator detectors provide additional background suppression over argon-based detectors.



Finally, backgrounds may arise from beam-produced particles or neutrino interactions outside the detector that produce activity inside the LArTPC.  These contributions are difficult to estimate with existing literature or the simulations described here, but are nonetheless important given the proximity of the examined locations to the major neutrino production sites.  Beam-related backgrounds can be determined with direct assays at each examined location with neutron and charged particle spectrometers, while $\nu$-produced 'dirt' backgrounds can be estimated by propagating the neutrino fluxes described here through a full simulation package employing neutrino interaction generators and subsequent propagation of all final state particles.  

\section{Conclusions}  
To examine the suitability of decay-in-flight neutrino beam facilities as sites of decay-at-rest neutrino physics experiments, we have investigated the largely-neglected contribution of low-energy neutrinos in the Fermilab NuMI beamline.  
We find overall levels of low-energy neutrino production at NuMI to be roughly equal per MW to those produced by more-studied spallation neutron facilities.  
Modest liquid argon detectors similar in size to the existing CAPTAIN or MicroBooNE experiments, if placed in the vicinity of the NuMI target or absorber, could yield thousands of decay-at-rest $\nu_e$ and $\nu_{\mu}$ detections.  
Moreover, a similar detector in a dedicated facility along a future Fermilab high-power long-baseline neutrino beam has the potential to yield copious decay-at-rest statistics valuable for properly interpreting $\nu_e$ data for supernova and CP-violation results from liquid argon TPCs.  
Detailed flux studies for future facilities, further investigation of DIF $\nu$ contamination, and dedicated background assays in NuMI near-absorber and near-target locations, are recommended as next steps in investigating the short- and long-term potential of this class of experiment.  

\vspace{1mm}
We would like to thank R. Hatcher and Z. Pavlovic for their support and assistance with the G4NuMI simulation.  We would also like to thank R. Svoboda and W. C. Louis for valuable discussions and comments on the manuscript.  This work was supported by Illinois Institute of Technology, Fermilab via its Intensity Frontier Fellowship Program, and by the Department of Energy National Nuclear Security Administration under Award Number DE-NA0000979 through the Nuclear Science and Security Consortium.     


\bibliography{numi}
\end{document}